\normalfont

\documentclass[10pt,conference,letterpaper]{IEEEtran}
\usepackage[pdftex]{graphicx}
\usepackage{graphicx}
\usepackage[table]{xcolor}
\usepackage{booktabs}
\usepackage{tabularx}
\usepackage{balance}
\usepackage{makecell}
\usepackage{array}
\usepackage{listings}

\def\fixme#1{\bgroup \color{red}{[{#1}]}\egroup}

\IEEEoverridecommandlockouts

\newcommand*\mycircle[1]{
	\raisebox{.5pt}{\textcircled{\raisebox{-.9pt} {#1}}}
}

\begin{document}

\bstctlcite{bstctl:etal}

\title{Open-Source GEMM Hardware Kernels Generator: Toward Numerically-Tailored Computations}

\author{\IEEEauthorblockN{Louis Ledoux\IEEEauthorrefmark{1}\IEEEauthorrefmark{2},
Marc Casas\IEEEauthorrefmark{1}\IEEEauthorrefmark{2}
}
\IEEEauthorblockA{\IEEEauthorrefmark{1}Barcelona Supercomputing Center, Barcelona, Spain}
\IEEEauthorblockA{\IEEEauthorrefmark{2}Universitat Polit\`ecnica de Catalunya, Barcelona, Spain}

E-mail: \{louis.ledoux, marc.casas\}@bsc.es\\

}

\maketitle


\begin{keywords}
GEMMs, matrix-matrix-multiply, full stack framework, automated pipeline, flopoco, OpenCAPI, OpenBLAS, High Performance Computing, approximate/trans/extended precision.
\end{keywords}

\vspace{-4mm}
\section{Extended Abstract}
\label{sec:Introduction}
Many scientific computing problems can be reduced to Matrix-Matrix Multiplications (MMM), making the General Matrix Multiply (GEMM) kernels in the Basic Linear Algebra Subroutine (BLAS) of interest to the high-performance computing community.
However, these workloads have a wide range of numerical requirements.
Ill-conditioned linear systems require high-precision arithmetic to ensure correct and reproducible results~\cite{bailey_high-precision_2009}.
In contrast, emerging workloads such as deep neural networks, which can have millions up to billions of parameters, have shown resilience to arithmetic tinkering~\cite{johnson_rethinking_2018} and precision lowering~\cite{courbariaux_binarized_2016}.

\begin{figure}[b!]
\vspace{-0.3cm}
  \centering
  \includegraphics[width=\columnwidth]{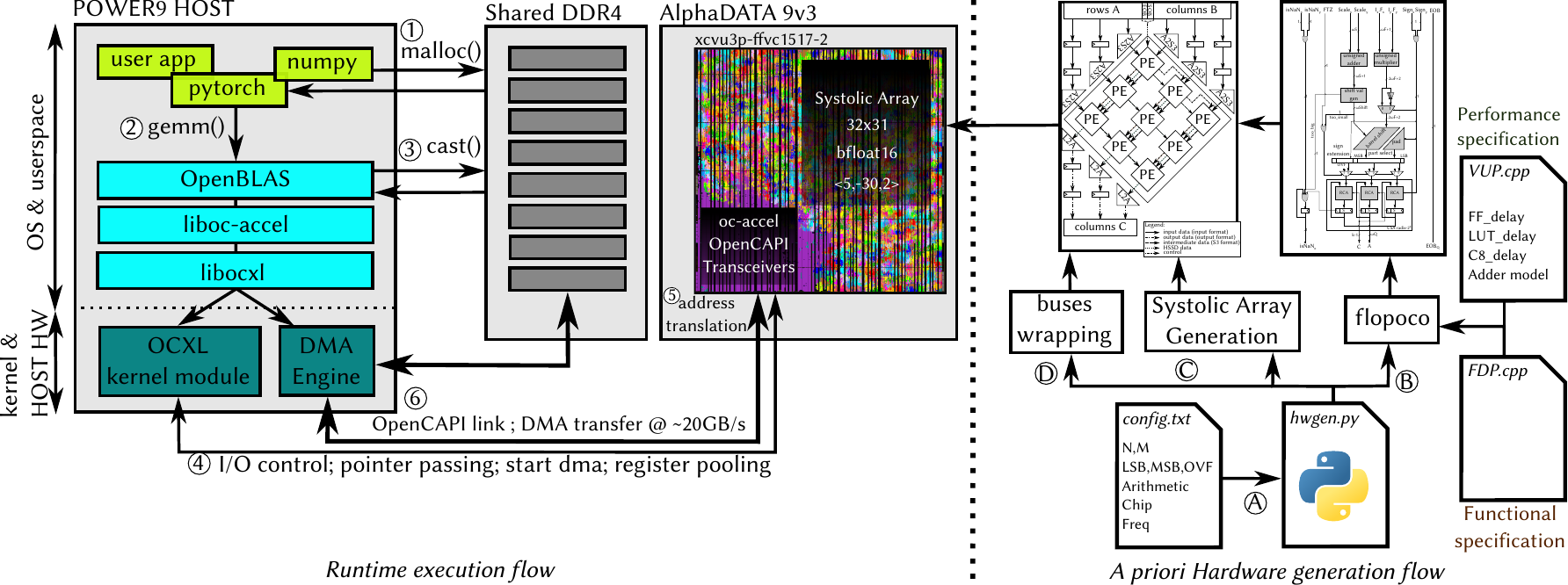}
  \vspace{-0.3cm}
	  \caption{Overview of the 2 phases framework. Left is Runtime execution flow and right is Hardware generation flow.}
  	\vspace{-0.3cm}
  \label{fig:overall_framework}
\end{figure}

General purpose arithmetic units and computer formats such as the IEEE754 standard naturally underperform in this vast land of scenarios.
We propose the generation of numerically tailored circuits where the necessary and sufficient internal precision is generated to target the computations requirements in terms of numerical quality while improving the energy cost.

\subsection{Open Source SW/HW co-designed framework for numerically tailored MMMs}
\label{sec:framework}
As depicted by Fig.~\ref{fig:overall_framework}, our framework~\cite{ledoux_generator_2022} is composed of two distinct phases, the \textit {a priori} Hardware generation flow and the runtime execution flow.
Because MMMs are  basically made of arbitrary long dot products, we design a custom Fused Dot Product (FDP) operator that is agnostic to the computer format and supports posit, IEEE754, and bfloat16 variations, while never rounding between two accumulations.
The intermediate precision of the fixed-point accumulator used in the dot-product is a key aspect of this work, and is configurable through the length of the scratchpad delimited by the parameters MSB (Most Significant Bit) and LSB (Least Significant Bit).
We leverage the automated pipeline feature of \emph{flopoco}~\cite{istoan_automating_2017}, which is an effective tool for efficiently exploring the wide range of functional specifications along with performance specifications, to produce MMM kernels with the necessary basic elements (LUTs, FFs, Carry chains, DSPs) for a targeted $(chip, frequency)$ couple (see Fig.~\ref{fig:overall_framework}-\mycircle{B}).

The essence of this work is to make intermediate precision tweakings from the hardware accessible to high-end software code as transparent as possible.
We achieve that by taking into account that many HPC codes rely on BLAS libraries to perform MMM operations.
Such libraries receive the function call to perform a GEMM and dispatch adequately to the underlying hardware at their disposal.

\begin{figure}[tb!]
  \centering
	\includegraphics[width=0.8\columnwidth]{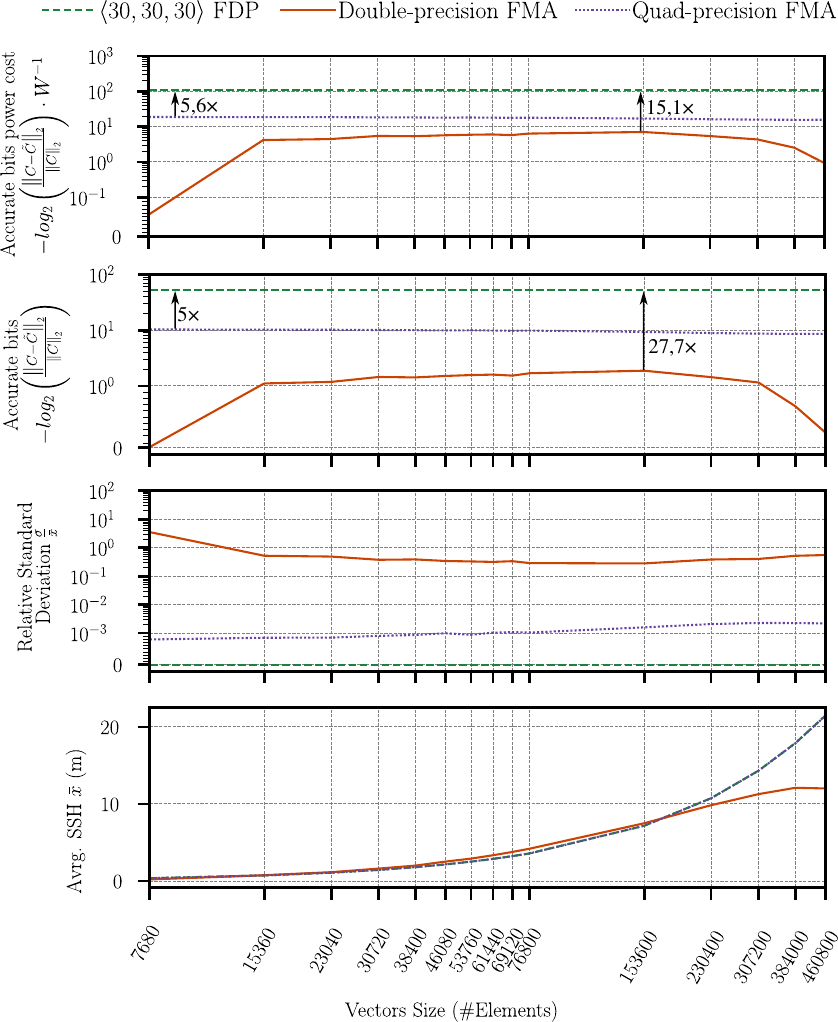}
  	\vspace{-0.3cm}
	  \caption{Sea Surface Height computation comparing IEEE-754 double-, quad-precision FMAs and a 91-bit FDP wrt numerical quality and power consumption.}
  	\vspace{-0.6cm}
  \label{fig:ssh_reproducibility}
\end{figure}

%
%
\begin{figure*}[th!]
  \centering
  \includegraphics[width=0.89\linewidth, height=7.8cm]{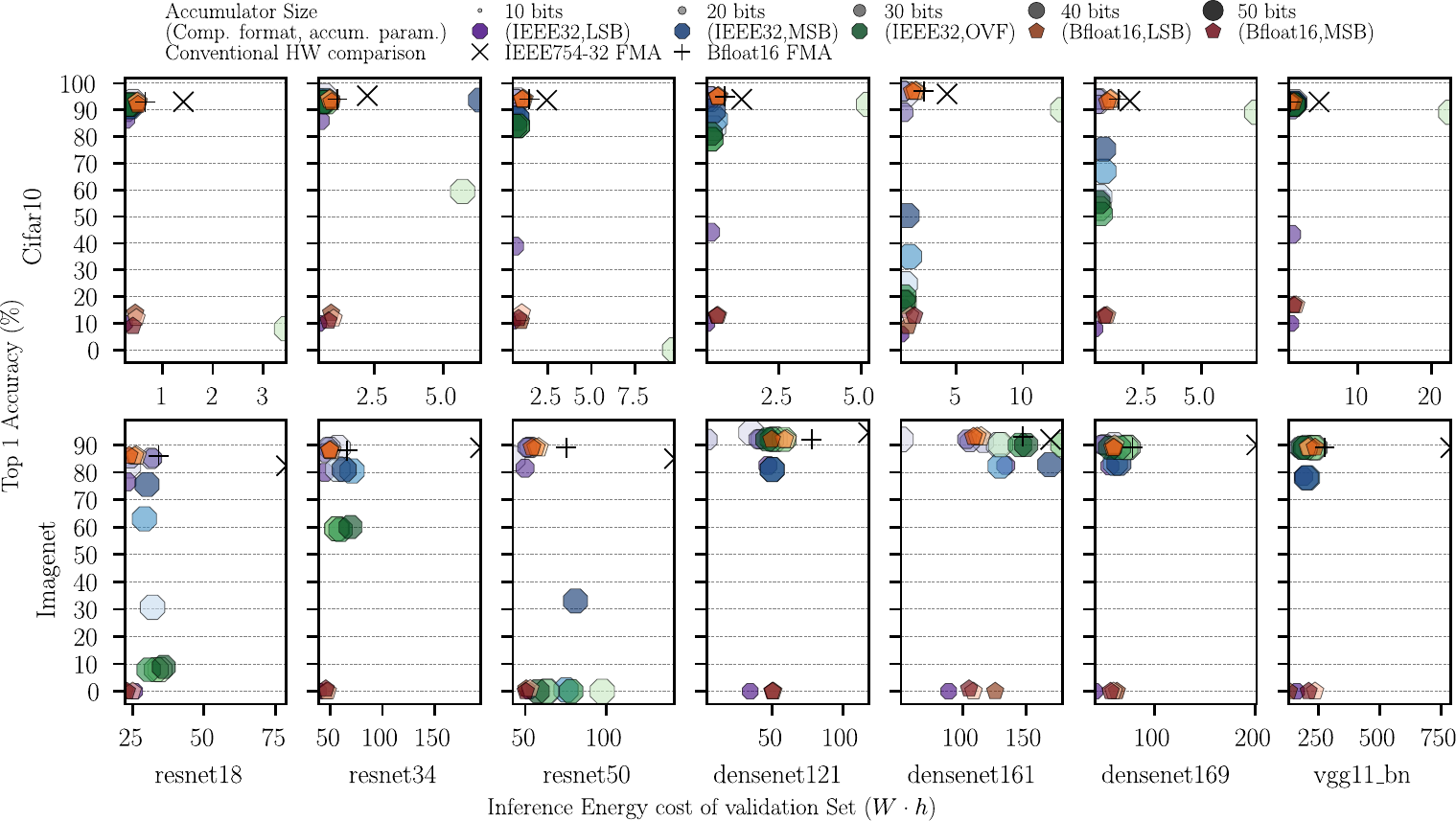}
  \vspace{-0.4cm}
	  \caption{Top1 Accuracy vs validation dataset inference energy cost for various combinations of datasets, models, computer formats, and accumulators.}
  	\vspace{-0.6cm}
  \label{fig:accuracy_vs_energy}
\end{figure*}

\vspace{-2mm}
\subsection{HPC workloads results}
\vspace{-2mm}
	We experiment with two families of real HPC workloads with contrasting numerical requirements, namely Artificial Intelligence (AI) and Sea Surface Height (SSH), whose respective results can be observed in Fig.~\ref{fig:accuracy_vs_energy} and Fig.~\ref{fig:ssh_reproducibility}.

For the SSH computation, the results obtained with 64-bit and 128-bit FPUs exhibit decreasing reproducibility as the vector size increases.
In contrast, our 91-bit $\left \langle ovf:30,msb:30,lsb:30 \right \rangle$ FDP maintains reproducibility for all vector sizes without deviation.
Our proposed FDP consistently exhibits 52 correct bits, which is at least 5$\times$ and 27.7$\times$ more than quad-precision and double-precision.
Our measurements on VU3P-2 FPGA at $200MHz$ show that the units power consumption are 0.266, 0.549, and 0.491 watts for double-precision FMA, quad-precision FMA, and the 91-bit FDP, respectively.
For all evaluated sizes, the 91-bit FDP yields at least $5.6\times$ and $15.1\times$ more correct bits for the same wattage as quad-precision and double-precision FMAs, respectively.

For AI workloads, we employ Pytorch as a base framework and link it to our modified OpenBLAS.
We use popular neural network models such as ResNet18, ResNet34, ResNet50, DenseNet121, DenseNet161, DenseNet169, and VGG11 with batch normalization, and evaluate them on the CIFAR-10 and ImageNet datasets.
To measure power consumption and accuracy, we use the BrainFloat16 and IEEE-754 32-bit formats for our computations with a large variety of accumulators varying their $OVF$, $MSB$, and $LSB$ parameters.
Fig.~\ref{fig:accuracy_vs_energy} shows the relationship between power consumption and accuracy for different accumulator and arithmetic combinations.
For example, if $84\%$ Top1 accuracy is satisfying for Imagenet with Resnet50, the most suited arithmetic/accumulator combination is IEEE-754 32-bit/$\left\langle ovf:9,msb:6,lsb:-20 \right\rangle$ represented by a light purple hexagon as all other markers are either on the right or below.\looseness=-1

\vspace{-2mm}
\subsection{Conclusion}
\vspace{-2mm}
Overall, our work highlights the importance of numerically tailored accumulators for reproducibility in scientific computing applications. Our results provide valuable insights into the trade-offs between power consumption and accuracy, and we believe that our results have the potential to inform the design of future AI and scientific computing systems. In light of this, we encourage other researchers to explore the possibilities of low precision accumulators using our open-source framework.


\IEEEpeerreviewmaketitle

\vspace{-2mm}

\section{Acknowledgment}
\vspace{-2mm}
Marc Casas is supported by Grant RYC-2017-23269 funded by MCIN/AEI/ 10.13039/501100011033 and by “ESF Investing in your future”.
\vspace{-4mm}

%
\balance
\bibliographystyle{IEEEtran}
\bibliography{IEEEabrv,references}
\vspace{-7mm}

\begin{biography}[
{
\includegraphics[width=1in,height=1.25in,clip,keepaspectratio]{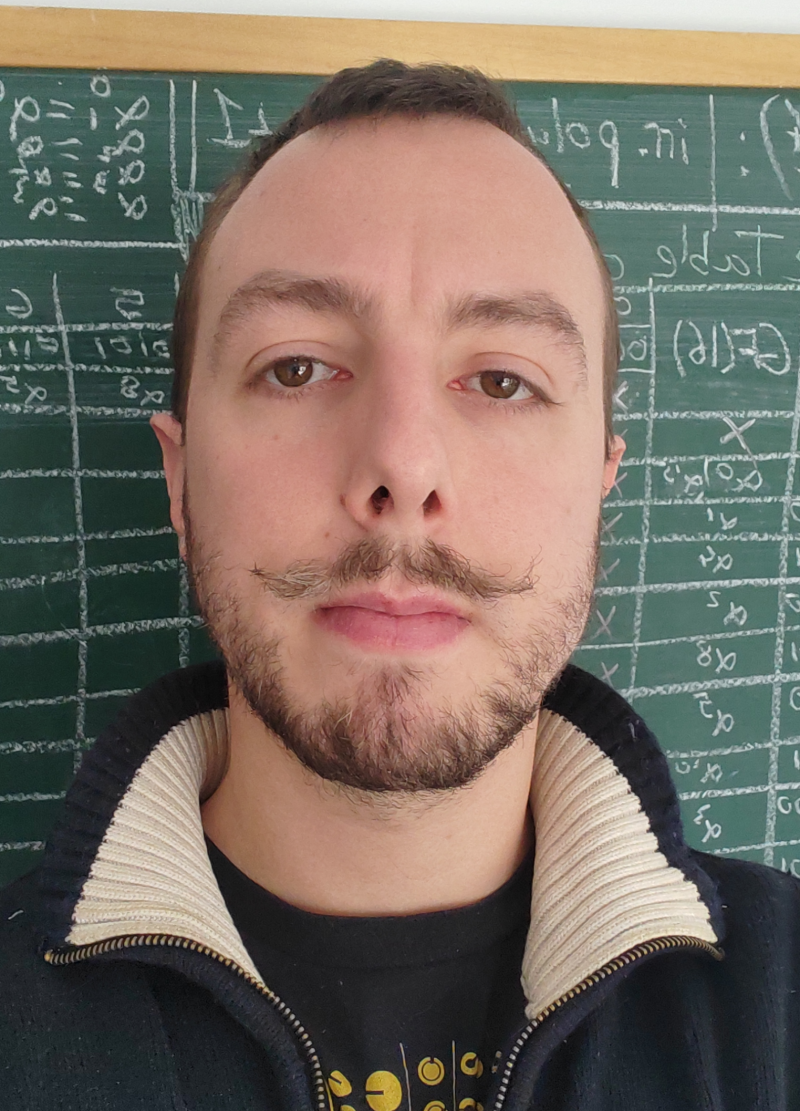}
}
]
	{Louis Ledoux} received his BSc degree in 2016 in Computer Science from Universit\'e de Rennes1, France. The following years, he pursued his MSc degree in parallel with an Engineer diploma from \'Ecole Sup\'erieure d'Ing\'enieurs de Rennes (ESIR). He concluded in 2018 his studies in Rennes with a position of Hardware Engineer at b\textless \textgreater com, a national research laboratory. This position allowed him to experiment with the first FPGAs in the cloud and their virtualizations. Since 2018, he has been a PhD candidate at the Computer Architecture departments of Barcelona Supercomputing Center (BSC) and Universitat Polit\`ecnica de Catalunya (UPC), Spain.

\end{biography}

\end{document}